\documentclass[english,floatfix,twocolumn,twoside,english,superscriptaddress,prl]{revtex4}
\usepackage[T1]{fontenc}
\usepackage[utf8]{inputenc}
\usepackage{amsmath}
\usepackage{graphicx}
\usepackage{color}

\definecolor{Red}{rgb}{1,0,0}
\definecolor{Blue}{rgb}{0,0,1}
\definecolor{Olive}{rgb}{0.41,0.55,0.13}
\definecolor{Green}{rgb}{0,1,0}
\definecolor{MGreen}{rgb}{0,0.8,0}
\definecolor{DGreen}{rgb}{0,0.55,0}
\definecolor{Yellow}{rgb}{1,1,0}
\definecolor{Cyan}{rgb}{0,1,1}
\definecolor{Magenta}{rgb}{1,0,1}
\definecolor{Orange}{rgb}{1,.5,0}
\definecolor{Violet}{rgb}{.5,0,.5}
\definecolor{Purple}{rgb}{.75,0,.25}
\definecolor{Brown}{rgb}{.75,.5,.25}
\definecolor{Grey}{rgb}{.5,.5,.5}
\definecolor{Pink}{rgb}{1,0,1}
\definecolor{DBrown}{rgb}{.5,.34,.16}

\definecolor{Black}{rgb}{0,0,0}

\makeatletter

\providecommand{\tabularnewline}{\\}

\makeatletter

\usepackage{hyperref}

\makeatother

\usepackage{babel}
\makeatother

\begin{document}

\title{Statistical Mechanics of Steiner trees}

\author{M. Bayati}

\affiliation{Microsoft Research, One Microsoft Way, 98052 Redmond, WA}

\author{C. Borgs}

\affiliation{Microsoft Research, One Microsoft Way, 98052 Redmond, WA}

\author{A. Braunstein }

\affiliation{Politecnico di Torino, Corso Duca degli Abruzzi 24, 10129 Torino,
Italy}

\author{J. Chayes}

\affiliation{Microsoft Research, One Microsoft Way, 98052 Redmond, WA}

\author{A. Ramezanpour}

\affiliation{ICTP, Strada Costiera 11, I-34100 Trieste, Italy}

\author{R. Zecchina}

\affiliation{Politecnico di Torino, Corso Duca degli Abruzzi 24, 10129 Torino,
Italy}

\begin{abstract}
The Minimum Weight Steiner Tree (MST) is an important combinatorial
optimization problem over networks that has applications in a wide
range of fields. Here we discuss a general technique to translate
the imposed global connectivity constrain into many local ones that
can be analyzed with cavity equation techniques. This approach leads
to a new optimization algorithm for MST and allows to analyze the
statistical mechanics properties of MST on random graphs of various
types.
\end{abstract}
\maketitle
Given a graph or a lattice, finding a subgraph that optimizes some
global cost function is an important problem in many fields. One of
the most basic versions of this is known as the Minimum Weight Steiner
Tree (MST) problem.

Given an undirected graph with positive weights on the edges, 
the MST problem consists in finding a connected
subgraph of minimum weight that contains a selected set of ``terminal''
vertices. Such construction may require the inclusion of some non\-terminal
nodes which are called Steiner nodes. Clearly, an optimal sub-graph
must be a tree. Solving MST is a key component of many optimization
problems involving real networks. Concrete examples are network reconstruction
in biology (phylogenetic trees and regulatory sub-networks), Internet
multi-casting, circuit design and power or water distribution networks
design, just to mention few famous ones. MST is also a beautiful mathematical
problem in itself which lies at the root of computer science being
both NP-complete \cite{Karp1972} and difficult to approximate \cite{robins2000ist}.
In  physics the Steiner tree problem has similarities with many
basic models such as polymers,  self avoiding walks or  transport 
networks (e.g. \cite{Durand2007})  with a non-trivial interplay between local an 
global frustration. 

Here we show that the cavity approach of statistical physics can be
used to both analyze and solve this problem on random graphs (as e.g.
\cite{Braunstein2005,MPZ,Braunstein2006b}) once an appropriate representation
is chosen. We actually study the even more general (and eventually
harder) $D-$MST problem in which we consider the depth of the tree
from a root terminal node to be bounded by $D$. Unfortunately the
traditional techniques for studying topologically connected structures,
as for instance the so-called $O(n)$ model, are incompatible
with the cavity method. We provide here instead an arborescent representation
of the Steiner problem which allows to implement explicitly global
connectivity constraints in terms of local ones.

In recent years many algorithmic results have appeared showing the
efficacy of the cavity approach for optimization and inference problems
defined over both sparse and dense random networks of constraints
\cite{MPZ,Braunstein2006b,Frey2007,Braunstein2003,Braunstein2005,di2004wdl}.
These performances are understood in terms of factorization properties
of the Gibbs measure over ground states, which can be also seen as
the onset of correlation decay along the iterations of the cavity
equations \cite{krzakal2007gsa}. 
Here we make a step further  by presenting evidence 
for the exactness of the cavity approach for a qualitatively different class of 
models, namely problems which are subject to rigid global constraints that couple all variables. 
Quite often this type of global constraint
is of topological origin and is common to many problems across disciplines
(e.g. the Traveling Salesman Problem in computer science or Self-Avioding
Walks in physics).

Our work addresses two questions: by analyzing the distributional
equations we provide the phase diagrams of the problem in the control
parameters $\alpha$ and $D$, where $\alpha N$ is the number of
terminals in a graph of $N$ vertices and $D$ is the allowed depth
of the tree from a randomly chosen root. We compute quantities like
the behavior of the minimum cost as a function of $D$ for a given
fraction $\alpha$ of terminals, or the number of Steiner nodes $cN^{s}$
where both $c$ and the exponent $s$ depend on $D$ and $\alpha$.
Such quantities are of extreme interest in that they are directly
connected with the topology of the tree. For instance, for the case
of complete graphs with random weights we find that an extremely small
depth $D_{N}$ is sufficient for reaching costs which are close to
optimal ones for the unbounded trees (e.g. for the complete graph
with random weights we find that $D_{N}\sim\log\log N$ is sufficient
to reach asymptotically a cost close to the optimal one  $\zeta(3)$ \cite{frieze1985vrm,gamarnik2005evr}
of the minimum spanning tree which has depth $\Theta(N^{1/3})$
\cite{breed:101}). For finite $D$ the results of the cavity approach
can be compared with rigorous upper and lower bounds \cite{Angel2008} making us conjecture
that the cavity approach is exact, as it happens for random Matchings
\cite{aldous2001zetalr}. 
Similar results hold for other classes of  random graphs. Here we give results for fixed 
degree and Scale-Free graphs, for which some non trivial patterns of solutions for optimal
Steiner trees appear.

On the algorithmic side, the arborescent representation of the problem
leads to cavity equations that can be turned into an algorithm for
solving single instances.

Very few results are known on the Steiner problem on random graphs
in the regime in which $\alpha$ is finite. For the complete graph
with random weights some upper and lower bounds for the minimum cost
have been derived \cite{Bollobas2004}, which are compatible with
those predicted by the cavity method. For finite degree random graphs
(e.g. Erdös-Rényi, fixed degree or scale-free graphs) much less is
known.

\paragraph{The model.}

We model the Steiner tree problem as a rooted tree (such a construction
is often associated with the term {}``arborescence''). Each node
$i$ is endowed with a pair of variables $\left(p_{i},d_{i}\right)$,
a pointer $p_{i}$ to some other node in the neighborhood $V(i)$
of $i$ and a depth $d_{i}\in\left\{ 1,\dots,D\right\} $ defined
as the distance from the root. Terminal nodes must point to some other
node in the final tree and hence $p_{i}\in\mathrm{V}(i)$. The root
node conventionally points to itself . Non-root nodes either point
to some other node in $\mathrm{V}(i)$ if they are part of the tree
(\emph{Steiner} and \emph{terminal} nodes) or just do not point to
any node if they are not part of the tree (allowed only for non-terminals),
a fact that we represent by allowing for an extra state for the pointer
$p_{i}\in V(i)\cup\emptyset$. The depth of the root is set to zero,
$d_{i}=0$ while for the other nodes in the tree the depths measure
the distance from the root along the unique oriented path from the
node to the root

In order to impose the global connectivity constraint for the tree
we need to impose the condition that if $p_{i}=j$ then 
$p_j\neq\emptyset$ and $d_{j}=d_{i}-1$.
This condition forbids loops and guarantees that the pointers describe
a tree. In building the cavity equations (or the Belief Propagation
equations), we need to introduce the characteristic functions $f_{ij}$
which impose such constraints over configurations of the independent
variables $\left(p_{i},d_{i}\right)$. For any edge $\left(i,j\right)$
we have the indicator function $f_{ij}=g_{ij}g_{ji}$ where $g_{jk}=\left(1-\delta_{p_{k},j}\left(1-\delta_{d_{j},d_{k}-1}\right)\right)\left(1-\delta_{p_{k},j}\delta_{p_{j},\emptyset}\right)$.

\paragraph{Cavity Equations.}

The cavity equations take the form \begin{align}
P_{j\to i}\left(d_{j},p_{j}\right) & \propto e^{-\beta c_{jp_{j}}}\prod_{k\in j\setminus i}Q_{k\to j}\left(d_{j},p_{j}\right)\label{eq:phat}\\
Q_{k\to j}\left(d_{j},p_{j}\right) & \propto\sum_{d_{k}p_{k}}P_{k\to j}\left(d_{k},p_{k}\right)f_{jk}\left(d_{k},p_{k},d_{j},p_{j}\right)\label{eq:qhat}\end{align}
 where $c_{ij}$ is the weight of the link $(i,j)$, with $c_{i\emptyset}=\infty$
if $i$ is a terminal. The $\propto$ symbol accounts for a multiplicative
normalization constant. Allowed configurations are weighted by $e^{-\beta c_{ij}}$
where $\beta^{-1}$ is a temperature fixing the energy level. The
zero temperature limit is taken by considering the following change
of variables: $\psi_{j\to i}\left(d_{j},p_{j}\right)=\beta^{-1}\log P_{j\to i}\left(d_{j},p_{j}\right)$
and $\phi_{k\to j}\left(d_{j},p_{j}\right)=\beta^{-1}\log Q_{k\to j}\left(d_{j},p_{j}\right)$.
In the $\beta\to\infty$ limit Eq.~\ref{eq:phat}-\ref{eq:qhat}
reduce to: \begin{align}
\psi_{j\to i}\left(d_{j},p_{j}\right)= & -c_{jp_{j}}+\sum_{k\in j\setminus i}\phi_{k\to j}\left(d_{j},p_{j}\right)\label{eq:psi}\\
\phi_{k\to j}\left(d_{j},p_{j}\right)= & \max_{d_k,p_k:f_{jk}\left(d_{k},p_{k},d_{j},p_{j}\right)\neq0}\psi_{k\to j}\left(d_{k},p_{k}\right)\label{eq:phi2}\end{align}

The previous two equalities must be understood to hold except for
an additive constant. Eqs.~\ref{eq:psi}-\ref{eq:phi2} are in the
so called \char`\"{}Max Sum\char`\"{} form. 

On a fixed point, one can compute \emph{marginals} $\psi_j$:
\begin{equation}
\psi_j\left(d_j,p_j\right)=-c_{jp_j}+\sum_{k\in j}\phi_{k\to j}(d_j,p_j)\label{eq:marg}\\
\end{equation}
and the optimum tree should be given by $\arg\max \psi_j$.

If the starting graph is a tree $\psi_{j\to i}(d_{j},p_{j})$
can be interpreted as the minimum cost change of removing a vertex
$j$ with forced configuration $d_{j},p_{j}$ from the subgraph with
link $\left(i,j\right)$ already removed. We introduce the variables
$A_{k\to j}^{d}\equiv\max_{p_{k}\neq j,\emptyset}\psi_{k\to j}\left(d,p_{k}\right)$,
$B_{k\to j}^{d}\equiv\psi_{k\to j}\left(d,\emptyset\right)$, $C_{k\to j}^{d}\equiv\psi_{k\to j}\left(d,j\right)$,
$D_{k\to j}=\max_{d}\max\{ A_{k\to j}^{d},B_{k\to j}^{d}\} $
and $E_{k\to j}^{d}=\max\{ C_{k\to j}^{d+1},D_{k\to j}\} $.
This is enough to compute $\phi_{k\to j}\left(d_{j},p_{j}\right)=A_{k\to j}^{d_{j}-1},D_{k\to j},E_{k\to j}^{d_{j}}$
for $p_{j}=k$, $p_{j}=\emptyset$ and $p_{j}\neq k,\emptyset$ respectively.
Eqs. \ref{eq:psi}-\ref{eq:phi2} can then be solved by repeated iteration
of the following set of equations: \begin{eqnarray}
A_{j\to i}^{d}(t+1) & = & \sum_{k\in j\setminus i}E_{k\to j}^{d}(t) + \label{eq:A}\\
& + & \max_{k\in j\setminus i}\{A_{k\to j}^{d-1}(t)-E_{k\to j}^{d}(t)-c_{jk}\} \nonumber \\
B_{j\to i}(t+1) & = & -c_{j\emptyset}+\sum_{k\in j\setminus i}D_{k\to j}(t)\label{eq:B}\\
C_{j\to i}^{d}(t+1) & = & -c_{ij}+\sum_{k\in j\setminus i}E_{k\to j}^{d}(t)\label{eq:C}\\
D_{j\to i}(t) & = & \max\left(\max_{d}A_{j\to i}^{d}\left(t\right),B_{j\to i}\left(t\right)\right)\label{eq:D}\\
E_{j\to i}^{d}(t) & = & \max\left(C_{j\to i}^{d+1}\left(t\right),D_{j\to i}\left(t\right)\right)\label{eq:E}\end{eqnarray}

For graphs without cycles the above equations are guaranteed to converge
to the optimal solution. In graphs with cycles, these equations may instead 
fail to converge in some cases. For the classes of random graphs studied 
in this work, this appears not to be due to a replica symmetry breaking 
instability but rather to the effect of local structures in the underlying 
graph (as it is
known to happen in simpler problems such as random matchings \cite{mezard1986mfe}).
This observation is corroborated by the analysis of the distributional
cavity equations discussed later. While more work is needed
to understand this point, from the algorithmic viewpoint the problem
can be overcome by applying a small perturbation \cite{Braunstein2006b}.
The term $\psi_j(d_j,p_j)$ of Eq.~\ref{eq:marg} multiplied by a (small) constant $\rho$ is
added to the rhs. of Eq.~\ref{eq:psi}. This leads to
a set of equations which show good convergence properties for vanishing $\rho$.

An equivalent formulation of the problem can be constructed by introducing
a link representation of the pointer variables (one may introduce
link variables $x_{ij}=0,\pm1$, $0$ if $i$ does not point $j$,
$1$ if $i$ points $j$ and $-1$ if $j$ points $i$). In this representation,
the number of states of the independent variables is just $3D$ which
can be kept finite for complete graphs or at most of order $\log N$
for sparse graphs.

\paragraph{Distributional equations and average case analysis.}

Population dynamics (or density evolution) is a powerful tool to solve
distributional equations that deal with a large number of random variables.
In the physics community the method was introduced in \cite{popdyn}
for the study of spin glass models on diluted random graphs. Population
dynamics is useful especially when the equations involve sums over
many states of the variables. The underlying idea is to represent
probability distributions with a population of random variables and
use the equations to update such populations. After a suitably large
number of updates the histogram of variables in the population will
converge to a stable distribution.

To obtain results on the $N\to\infty$ limit one would need to rescale
simultaneously all $d$-dependent quantities in order to eliminate
their direct dependence on $N$ in Eqs.~\ref{eq:A}-\ref{eq:E}.
We limited however ourselves here for all cases analyzed to large
but finite $N$, in particular because the obviously needed dependence
of $D$ on $N$ for finite degree graphs makes this task even more
involved.

We will apply the population dynamics method to find the statistical
properties of the cavity fields $M_{i\to j}=\left(A_{i\rightarrow j}^{d},B_{i\rightarrow j},C_{i\rightarrow j}^{d},D_{i\rightarrow j},E_{i\rightarrow j}^{d}\right)$
in Eqs. \ref{eq:A}-\ref{eq:E}. Given an ensemble of random graphs
we will find the probability distribution of these fields from which
we will derive the quantities of interest, namely the average minimum
cost and average number of Steiner nodes as a function of $N$, in
the so called Bethe approximation which is implicit in the cavity
approach. The method proceeds by initializing at random a population 
of field vectors $M_{a}=\left(A_{a}^{d},B_{a},C_{a}^{d},D_{a},E_{a}^{d}\right)$ with $a\in[0,N_{p}]$ and $d\in[0,D]$. The first member $M_0$
represents messages sent by root. Members
with label $a=1,\dots,N_{t}$ represent messages sent by terminal nodes.
Here $N_{t}=\alpha N_{p}$ where $\alpha=K/N$
is the fraction of terminal nodes. Then the population dynamics algorithm
works by updating the population using Eqs. \ref{eq:A}-\ref{eq:E}
until convergence is reached. For brevity, we omit the details of
this procedure. Once convergence is reached, marginals 
$\psi_{a}\left(d,p\right)$ can be computed using Eq.~\ref{eq:marg}. 
The state $\left(d^{*},p^{*}\right)$
that maximizes the local marginal gives the energy contribution of 
the $a-th$ member.
If $p^{*}\neq\emptyset$ and $N_t<a$, then $a$ is a Steiner member. Finally the minimum cost reads
$E=Ke_{t}+\left(N-K\right)e_{s}$ where $e_{t}$ and $e_{s}$ are
the average energy of terminal and Steiner members.
The fraction of Steiner members in the population will give the fraction
of Steiner nodes in the ensemble of random graphs.

\begin{figure}
\includegraphics[width=1\columnwidth]{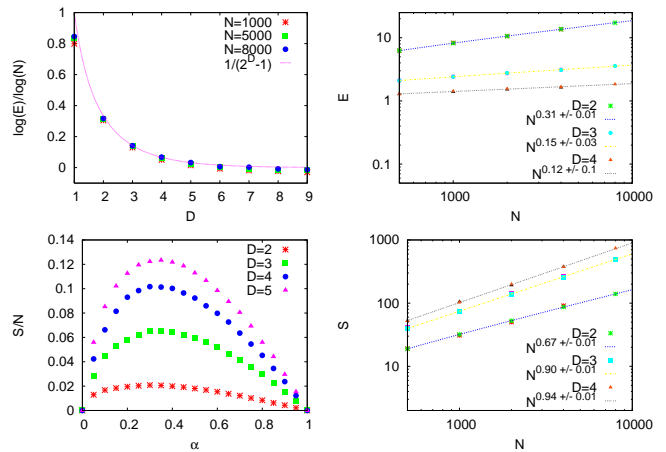}

\caption{$D$-MST on complete graphs. Left: Minimum cost (at $\alpha=0.5$)
and fraction of Steiner nodes (for $N=8000$) as a function of $D$.
Right: Comparison of Pop. dyn. with the algorithm on single samples
for various values of $N$ at $\alpha=0.5$. Fits are in very good
agreement with known bounds.}

\label{fig:complete-graph}
\end{figure}

\begin{figure}
\includegraphics[width=1\columnwidth]{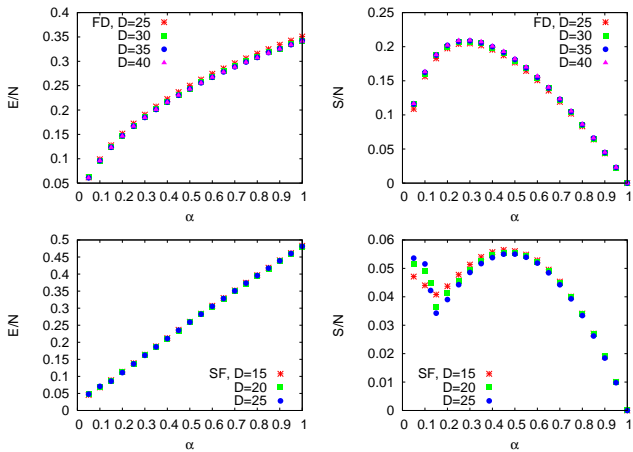}

\caption{Fixed degree (FD) and scale-free (SF) graphs. Left: Minimum cost as
function of $\alpha$ for different values of $D$. Right: Fraction
of Steiner nodes as a function of $\alpha$. The FD graphs have degree
$C=3$ and size $N=10^{6}$. The SF graphs have exponent $\gamma=3$
and size $N=10^{4}$.}

\label{fig:fix-degree-sf}
\end{figure}

\begin{figure}
\includegraphics[width=1\columnwidth]{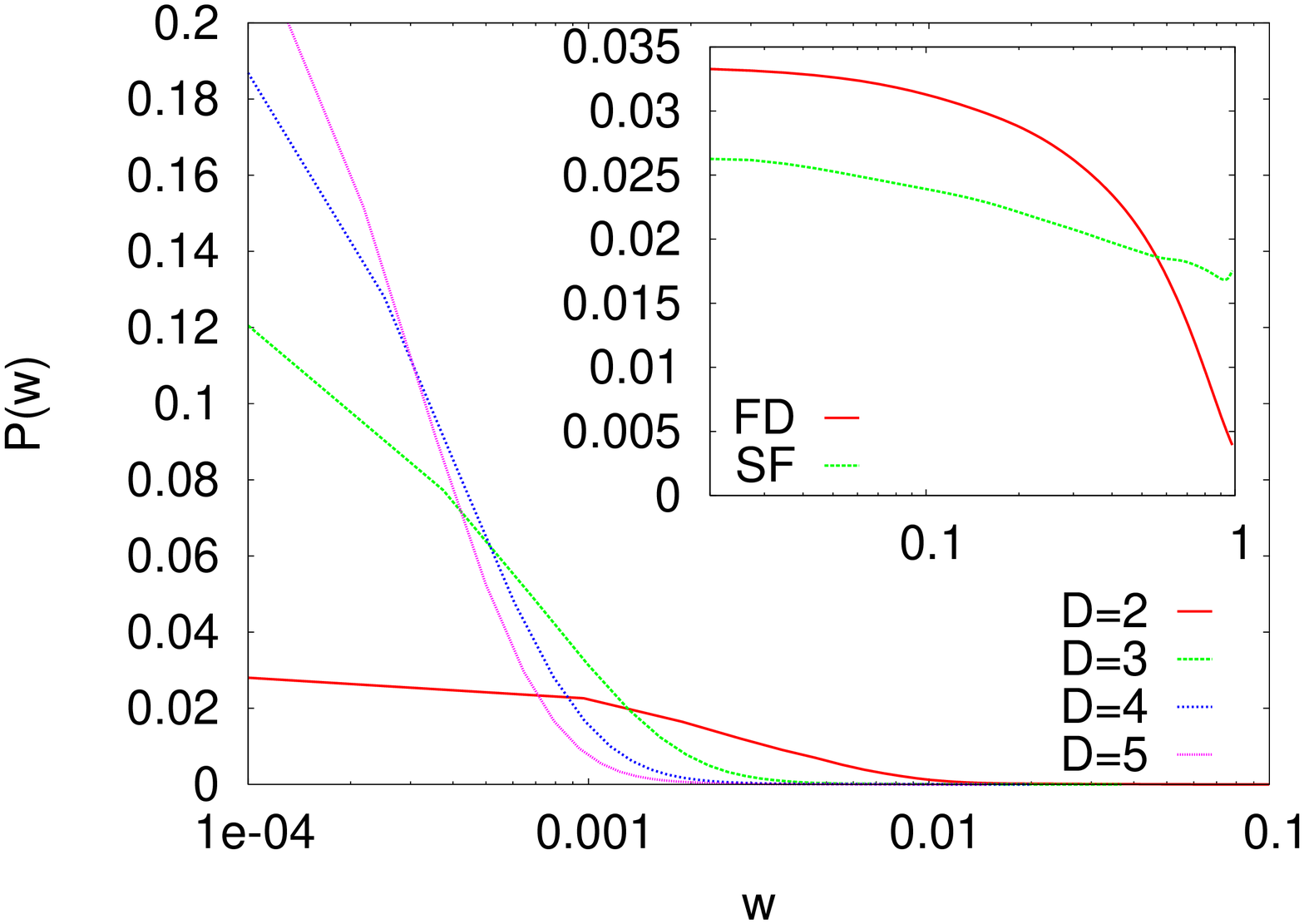}

\caption{Weight distribution of the MST for complete graphs of size $N=8000$
at $\alpha=0.5$. Inset: For FD graphs of degree $C=3$ ($N=10^{6}$)
and SF graphs of exponent $\gamma=3$ ($N=10^{4}$) with parameters
$D=25$, $\alpha=0.5$.}

\label{fig:distributions}
\end{figure}

In Figures \ref{fig:complete-graph}-\ref{fig:distributions} we display
numerical results for three classes of random graphs, namely complete
graphs, finite connectivity random graphs and scale-free graphs. We
first verify a quite remarkable agreement between the output of the
algorithm which finds Steiner trees on given random instances with
the outcomes of the population dynamics averaged over the randomness.
In Figs 1-2, we estimate the dependence on the depth $D$ of the minimum
cost and of the size of the Steiner set nodes. For complete graph
with random weights we are able to provide an accurate estimate of
the scaling exponents which for $\alpha=1$ are compatible
with rational exponents predicted by rigorous analysis \cite{Angel2008}.
Moreover, we observe a very rapid decrease of the minimum cost with
$D$, compatible with $N^{1/\left(2^{D}-1\right)}$. This suggests
that very few \char`\"{}hops\char`\"{} ($\sim\log\log N$) are indeed
sufficient to reach optimal costs. From a qualitative point of view
we observe a non trivial dependence on $N$ and $\alpha$ of the size
of the Steiner set. The size itself turns out to be sublinear, with
a rational exponent that depends on $D$. For fixed $N$ there appears
a maximum for relatively small values of $\alpha$. For the Scale-Free
graphs there appears an additional cuspid-like minimum. Finally, in
Fig.~\ref{fig:distributions} we provide the probability distribution of optimal weights
for all classes.

We conclude this letter by mentioning the connection with rigorous
results. For the case of bounded depth trees on complete graphs our numerical results show that the cavity equations
are indeed consistent with known bounds. As discussed in \cite{Angel2008}, 
the analysis of a simple greedy algorithm and a Chernoff-type bound lead to upper
and lower bounds for the minimum cost that are able to identify the
exact scaling exponent and to give bounds for the pre-factors. More
precisely, it can be shown that the average minimum $E_{D}$ grows with the size  as $N^{1/(2^{D}-1)}$ .
The case $D=2$ and $\alpha=1$ is particularly easy to understand:
the greedy algorithm amounts at choosing a first set of $N_{1}$ nodes
at depth $1$ by selecting the $N_{1}$ links with smallest weights.
Successively the remaining $N-N_{1}$ nodes at depth $2$ are connected
to the first layer by choosing the smallest weight for each node.
By optimizing over the size of $N_{1}$ one finds for the average
minimum cost $E_2=\frac{3}{2}N^{1/3}$ (a naive guess may give an
exponent $1/2$ instead of $1/3$). Comparisons with the cavity approach
for small $D$ show that indeed the exponent is $1/\left(2^{D}-1\right)$
as it should and that there exist a constant additional (negative)
term to the minimum cost which improves over the greedy algorithm.
Table \ref{table} shows the results of a power law fit to our data
for the average minimum cost and number of Steiner nodes as a function
of $N$. 
For  $D=N-1$ and $\alpha=1$ it is possible to prove using
techniques based on the computation tree that if the BP equations
converge, then the result is optimal. Details about these results
and hopefully about their extensions to the $\alpha<1$ case will
be given elsewhere. Work is in progress  to apply the algorithmic scheme we  have presented   to clustering,  network reconstruction and protein pathways  identification problems.

\begin{table}
\begin{centering}
\begin{tabular}{c|c|c|rcl|rcl|rcl}
 & $D$  & $\alpha$  &  & $a$  &  &  & $b$  &  &  & $c$  & \tabularnewline
\hline
$E$  & $2$  & $0.5$  & $-1.07$  & $\pm$  & $0.07$  & $0.92$  & $\pm$  & $0.01$  & $0.31$  & $\pm$  & $0.01$ \tabularnewline
$S$  & $2$  & $0.5$  & $-3.62$  & $\pm$  & $0.13$  & $0.35$  & $\pm$  & $0.01$  & $0.67$  & $\pm$  & $0.01$ \tabularnewline
$E$  & $3$  & $0.5$  & $-0.83$  & $\pm$  & $0.05$  & $1.21$  & $\pm$  & $0.02$  & $0.15$  & $\pm$  & $0.03$ \tabularnewline
$S$  & $3$  & $0.5$  & $0$  &  &  & $0.14$  & $\pm$  & $0.01$  & $0.90$  & $\pm$  & $0.01$ \tabularnewline
$E$  & $2$  & $1$  & $-1.46$  & $\pm$  & $0.25$  & $1.47$  & $\pm$  & $0.03(3/2)$  & $0.35$  & $\pm$  & $0.01(1/3)$ \tabularnewline
$E$  & $3$  & $1$  & $-0.95$  & $\pm$  & $0.05$  & $1.75$  & $\pm$  & $0.02$  & $0.15$  & $\pm$  & $0.02(1/7)$ \tabularnewline
\end{tabular}
\par\end{centering}

\caption{Comparing the exponents and prefactors for complete graphs. The parameters
have been obtained by fitting data to $a+bx^{c}$. In all the data
$N\le8000$. Values in the parenthesis are known analytical results.}

\label{table}
\end{table}

\end{document}